\documentclass[reprint,aps,prb,a4,superscriptaddress]{revtex4-1}

\usepackage{graphicx} 
\usepackage{dcolumn,amsmath,amssymb,amsfonts}
\usepackage{color} 
\usepackage{bm}

\begin{document}

\title{First principles-based calculation of the electrocaloric effect
  in BaTiO$_3$: comparison between direct and indirect methods}

\author{Madhura Marathe} \email{madhura.marathe@mat.ethz.ch}
\affiliation{Materials Theory, ETH Z\"urich, Wolfgang-Pauli-Str. 27,
  8093 Z\"urich, Switzerland}

\author{Anna Gr\"unebohm} \affiliation{Faculty of Physics and CENIDE,
  University of Duisburg-Essen, 47048, Duisburg, Germany}

\author{Takeshi Nishimatsu} \affiliation{Institute for Materials
  Research, Tohoku University, Sendai 980-8577, Japan}

\author{Peter Entel} \affiliation{Faculty of Physics and CENIDE,
  University of Duisburg-Essen, 47048, Duisburg, Germany}

\author{Claude Ederer} \email{claude.ederer@mat.ethz.ch}
\affiliation{Materials Theory, ETH Z\"urich, Wolfgang-Pauli-Str. 27,
  8093 Z\"urich, Switzerland}

\date{\today}

\begin{abstract}
We use molecular dynamics simulations for a first principles-based
effective Hamiltonian to calculate two important quantities
characterizing the electrocaloric effect in BaTiO$_3$, the adiabatic
temperature change $\Delta T$ and the isothermal entropy change
$\Delta S$, for different electric field strengths. We compare direct
and indirect methods to obtain $\Delta T$ and $\Delta S$, and we
confirm that both methods indeed lead to identical result provided
that the system does not actually undergo a first order phase
transition. We also show that a large electrocaloric response is
obtained for electric fields beyond the critical field strength for
the first order phase transition. Furthermore, our work fills several
gaps regarding the application of the first principles-based effective
Hamiltonian approach, which represents a very attractive and powerful
method for the quantitative prediction of electrocaloric
properties. In particular, we discuss the importance of maintaining
thermal equilibrium during the field ramping when calculating $\Delta
T$ using the direct method within a molecular dynamics approach.
\end{abstract}

\maketitle

\section{Introduction}

The ongoing search for alternative cooling technologies which are more
energy-efficient and environmentally friendly than conventional
vapor-compression refrigerators and offer the additional possibility
for device miniaturization has boosted research activities within the
fields of electrocaloric, elastocaloric, and magnetocaloric
effects.~\cite{Faehler_et_al:2011,Manosa/Planes/Acet:2013,Moya/Narayan/Mathur:2014}
The common feature in all three cases is that the application of an
external field (either electric, stress, or magnetic field) under
adiabatic conditions, i.e.\ when the active material is thermally
isolated from the environment, results in a temperature change of the
corresponding material. This reversible temperature change can be used
to transfer heat from a cool reservoir (the heat load) to a warmer
reservoir (e.g.\ the environment), thereby lowering the temperature of
the heat load (or keeping it at constant low temperature). It has been
found that such caloric effects are especially large close to ferroic
first order phase transitions, where giant responses can be triggered
through relatively modest
fields.\cite{Faehler_et_al:2011,Manosa/Planes/Acet:2013,Moya/Narayan/Mathur:2014}

In particular the electrocaloric (EC) effect has become very
attractive for potential future applications, due to the discovery of
a giant EC temperature change of 12\,K in Pb(Zr,Ti)O$_3$ thin
films.~\cite{Mischenko_et_al_2006} Here, the high crystalline quality
that can be achieved in thin film samples allows the application of
rather high electric fields without triggering a dielectric breakdown
of the samples. In recent years, a large number of studies -- both
theoretical and experimental -- have contributed to a better
understanding of the EC effect (see, e.g.
Refs.~\onlinecite{Scott_2011,Valant_2012,ECE-Book} and references
therein).

Nevertheless, direct measurements of the adiabatic temperature change
are still rather challenging, in particular for the case of thin film
samples.
Therefore, an indirect determination of this temperature change is
often preferred. The indirect method is based on a thermodynamic
Maxwell relation connecting the isothermal field-induced entropy
change with the temperature dependence of the electric polarization at
fixed electric field:
\begin{equation}
\left.\left(\frac{\partial S}{\partial \mathcal{E}}\right)\right|_T =
\left.\left(\frac{\partial P}{\partial T}\right)\right|_\mathcal{E} \ .
\label{eq:maxwell}
\end{equation}
The adiabatic temperature change $\Delta T$ can then be obtained from
pyroelectric measurements, i.e.\ by measuring the electric polarization
$P$ as function of temperature $T$ at different electric fields
$\mathcal{E}$:
\begin{equation}\label{eq:deltaT-ind}
\Delta T = -\int_{\mathcal{E}_1}^{\mathcal{E}_2}
\frac{T}{C_{p,\mathcal{E}}} \left.\left(\frac{\partial P}{\partial
  T}\right)\right|_\mathcal{E} \text{d}\mathcal{E}\ .
\end{equation}
Here, $C_{p,\mathcal{E}}$ is the specific heat at constant pressure
and applied field, and the external field is varied from
$\mathcal{E}_1$ to $\mathcal{E}_2$. It has to be noted that, if the
system undergoes a first order phase transition, the derivative
$\partial P/\partial T$ is ill-defined and the specific heat diverges,
which in principle does not allow application of
Eq.~\eqref{eq:deltaT-ind}. Furthermore, a possible contribution to the
EC effect stemming from the latent heat of the first order phase
transition is not accounted for by Eq.~\eqref{eq:deltaT-ind}.
Instead, the Clausius-Clapeyron equation has to be used to obtain the
corresponding contribution. In addition, the indirect method is only
suitable for ergodic systems. For example, it was shown that the
results from direct and indirect measurements do not match for relaxor
polymers,~\cite{Lu_et_al:2010} but compare well for ``normal''
ferroelectric polymers,~\cite{Lu_et_al:2011} Finally, the influence of
domains and anisotropy effects are not covered by the scalar form of
the Maxwell relation, Eq.~\eqref{eq:maxwell}.~\cite{Niemann2014281}

Another important quantity for characterizing the EC effect is the
isothermal entropy change $\Delta S$, which is related to the amount
of heat that is required to keep the system at constant temperature
while an electric field is applied or removed. The isothermal entropy
change can also be obtained indirectly from pyroelectric measurements
by simply integrating Eq.~\eqref{eq:maxwell}:
\begin{equation}
\Delta S = \int_{\mathcal{E}_1}^{\mathcal{E}_2}
\left.\left(\frac{\partial P}{\partial T}\right)\right|_\mathcal{E}
\text{d}\mathcal{E}\ .
\label{eq:deltaS-ind}
\end{equation}
On the other hand, $\Delta S$ can also be obtained in a (quasi-)
direct way from integrating the specific heat at constant electric
field:
\begin{equation}
\Delta S = \int_{T_1}^{T} \frac{C_{p,\mathcal{E}_1} -
  C_{p,\mathcal{E}_2}}{T'}\text{d}T'\ .
\label{eq:deltaS-dir}
\end{equation}
Note that strictly speaking this relation is only valid for $T_1
\rightarrow 0$. Nevertheless, for sufficiently low $T_1$ one can
assume that $S(T_1,\mathcal{E}_1) \approx S(T_1,\mathcal{E}_2)$ and
then Eq.~\eqref{eq:deltaS-dir} can be expected to give a good estimate
of $\Delta S$.\cite{Bai_et_al_2012,Moya/Narayan/Mathur:2014}

In the work presented in this article, we use a first principles-based
effective Hamiltonian
approach~\cite{Zhong_Vanderbilt_Rabe_1994,Zhong_Vanderbilt_Rabe_1995,Nishimatsu_et_al_2008}
to calculate the EC effect in the prototypical ferroelectric perovskite
BaTiO$_3$, and to address the applicability of the indirect method for
evaluating $\Delta T$ and $\Delta S$. Performing micro-canonical
molecular dynamics (MD) on the effective Hamiltonian allows for a
direct calculation of the EC temperature change under application or
removal of an electric field. Within the same framework, the
temperature dependence of the electric polarization under different
electric fields can be calculated and the temperature and entropy
changes can then be evaluated via Eqs.~\eqref{eq:deltaT-ind} and
\eqref{eq:deltaS-ind}. Thus, the effective Hamiltonian provides a
simplified but nevertheless realistic ``testing ground'' for the
general applicability of the indirect methods.

Previous studies employing first principles-based effective
Hamiltonians have found good agreement between direct and indirect
calculations of the EC temperature
change,\cite{Ponomareva/Lisenkov:2012,Nishimatsu_Barr_Beckman:2013}
provided that both $\mathcal{E}_1$ and $\mathcal{E}_2$ are above the
critical field for the first order phase transition, i.e.\ in a regime
where no discontinuities of the polarization occur as function of
temperature and electric field.  In
Ref.~\onlinecite{Ponomareva/Lisenkov:2012} the EC temperature change
for Ba$_{0.5}$Sr$_{0.5}$TiO$_3$ has been calculated using
micro-canonical Monte Carlo simulations (Creutz algorithm), and the
so-obtained values have been compared with the indirect evaluation
based on Eq.~\eqref{eq:deltaT-ind}, where $P(T,\mathcal{E})$ has been
obtained from standard Monte Carlo simulations within the canonical
ensemble. In Ref.~\onlinecite{Nishimatsu_Barr_Beckman:2013}, the
direct calculation of $\Delta T$ for BaTiO$_3$ has been performed
using a micro-canonical MD algorithm. The indirect evaluation of
$\Delta T$ using Eq.~\eqref{eq:deltaT-ind} showed reasonable agreement
with the corresponding directly calculated values. Discrepancies were
attributed to inconsistencies arising from an empirical,
temperature-dependent, pressure correction and to the use of a
constant empirical value for the specific heat (the experimental value
for $C_{p,\mathcal{E}}$ at room temperature was used in
Ref.~\onlinecite{Nishimatsu_Barr_Beckman:2013}). It is important to
note that $C_{p,\mathcal{E}}$ is not constant and varies significantly
with temperature and applied field, especially near the phase
transition.~\cite{Novak}

Here, we calculate the specific heat of the effective Hamiltonian, as
function of temperature and electric field, in order to allow for a
fully consistent comparison between the direct and indirect evaluation
of $\Delta T$ and $\Delta S$. We confirm that both methods indeed lead
to identical result provided that the system does not actually undergo
a first order phase transition. We also show that the actual
transition is not crucial for obtaining a sizable EC response and
compare this with the case of magnetocaloric Heusler
alloys. Furthermore, we calculate the isothermal EC entropy change and
again demonstrate good agreement between direct and indirect
methods. Finally, we investigate how fast the electric field can be
changed within the MD simulation without the system going out of
thermal equilibrium. In particular, we demonstrate the importance of
maintaining equilibrium during the simulation by monitoring changes in
the total energy of the system.

This paper is organized as follows. In Section~\ref{sec:theory}, we
briefly describe our computational method. Our results are then
presented in Section~\ref{sec:results}, which is divided into two
parts, the first describing the effect of different ramping rates for
the electric field, the second discussing the EC temperature and
entropy changes. Finally, in the last section, we summarize our main
results and conclusions.

\section{Computational method}
\label{sec:theory}

For our study, we use the effective Hamiltonian proposed by Zhong
\textit{et
  al.}.~\cite{Zhong_Vanderbilt_Rabe_1994,Zhong_Vanderbilt_Rabe_1995}
This effective Hamiltonian is applicable to ferroelectrics with a
cubic perovskite parent structure. The ferroelectric polarization in
these materials can be described by a relative displacement of cations
and anions, represented by a soft mode variable in the Hamiltonian.
In addition, local strain variables are included. This type of
description retains the dominant terms in the total energy while
reducing the number of degrees of freedom per unit cell from 15 to 6
(3 soft mode variables and 3 local strain variables). 

All parameters for the effective Hamiltonian can be obtained using
\textit{ab initio} density functional theory
calculations.~\cite{Zhong_Vanderbilt_Rabe_1995,Nishimatsu_et_al_2010}
The effective Hamiltonian approach is therefore able to determine
temperature-dependent properties of ferroelectric materials without
the need for empirical input parameters. For example, it was
demonstrated that the three consecutive phase transitions in bulk
BaTiO$_3$ are successfully
reproduced.~\cite{Zhong_Vanderbilt_Rabe_1994} Furthermore, the
effective Hamiltonian approach has been used successfully for the
calculation of EC properties.~\cite{
  Prosandeev/Ponomareva/Bellaiche:2008,Lisenkov/Ponomareva:2009,Ponomareva/Lisenkov:2012,Beckman_et_al_2012,Nishimatsu_Barr_Beckman:2013,Marathe/Ederer:2014}

We perform MD simulations employing the effective Hamiltonian as
implemented in the feram code~\cite{Nishimatsu_et_al_2008}
(\url{http://loto.sourceforge.net/feram/}), using the available
parameter set for BaTiO$_3$,~\cite{Nishimatsu_et_al_2010} which has
been obtained using the generalized gradient approximation for the
exchange-correlation functional according to Wu and
Cohen.~\cite{Wu/Cohen:2006}

In order to \emph{directly} calculate the adiabatic EC temperature
change, we first thermalize the system at a given temperature and
electric field using a Nos\'e-Poincar\'e
thermostat.~\cite{Bond/Leimkuhler/Laird:1999} We then switch off the
thermostat, i.e.\  we switch to the micro-canonical ensemble, and
slowly change the electric field while monitoring the resulting
changes in the total and kinetic energies. These calculations are
performed using a $96 \times 96 \times 96$ supercell,
i.e.\ corresponding to 96 simple perovskite unit cells along each
cartesian direction. A time step of 1\,fs per MD step is used and the
thermalization (averaging) time for these direct calculations is equal
to 80\,ps (40\,ps).
As usual, the temperature is calculated from the kinetic energy,
$E_\text{kin}$, of the system:
\begin{equation}
T = \frac{2E_\text{kin}}{N_f k_B} \ ,
\end{equation}
where $N_f$ denotes the number of degrees of freedom of the system and
$k_B$ is the Boltzmann constant. The EC temperature change $\Delta T$
is then simply obtained from the difference between the initial and
final temperature of the system, i.e.\ before and after the electric
field is ramped on or off. 

To reduce the computational effort, we use a simplified treatment for
the local strain variables, which are obtained by minimization of the
total energy for the current soft-mode configuration in each MD
step. Thus, our model contains only $N_f=3$ dynamic degrees of freedom
per unit cell (the 3 soft mode variables), compared to the original
15. As a result, the model specific heat and the directly calculated
$\Delta T$ need to be rescaled before comparing to experimental
data.~\cite{Nishimatsu_Barr_Beckman:2013} However, since the focus of
this work is on the internal consistency within the model description,
in order to assess the general validity of the indirect determination
of $\Delta T$ and $\Delta S$, and not on a quantitative comparison
with experimental data, we do not perform such rescaling within this
work, i.e.\ except where otherwise noted, all presented values for
$\Delta T$ and $C_{p,\mathcal{E}}$ refer to the model system and not
to the real material. We also note that a simple rescaling of $\Delta
T$ neglects the fact that the simulation corresponds to a ``wrong''
final state of the system, i.e.\ with different temperature and
polarization compared to the final state that would be obtained in a
real experiment, and therefore corrects only partially for the missing
degrees of freedom.

To calculate the adiabatic temperature change and the isothermal
entropy change using the \emph{indirect} method, we calculate
polarization as function of temperature (on a 1\,K grid) at several
applied electric fields using a $16 \times 16 \times 16$ simulation
cell, a thermalization time of 120\,ps and an averaging time of
80\,ps, with a 2\,fs time step per MD iteration. These calculations
are performed in the canonical ensemble using the Nos\'e-Poincar\'e
thermostat. We then use smoothing cubic spline functions to fit the
polarization versus temperature data, in order to determine $\left(
\partial P/\partial T\right)_\mathcal{E}$. 

The specific heat of the model Hamiltonian at constant pressure and
electric field, required for the indirect calculation of $\Delta T$
and the (quasi-) direct calculation of $\Delta S$, is determined by
calculating the derivative of the total energy, i.e.\  by using the
relation $C_{p,\mathcal{E}} = \left( \partial E_{tot}/\partial T
\right)_{p,\mathcal{E}}$, which is applicable for our simulations
performed at zero pressure. To calculate $E_{tot}(T)$, we use a
96$\times$96$\times$96 simulation cell, equilibration and averaging
times of 80\,ps and 40\,ps, respectively, and a 2\,fs time step. The
temperature dependence of $C_{p,\mathcal{E}}$ has been calculated
using ``cooling'' as well as ``heating'' simulations, i.e.\ where the
system at a particular temperature is initialized from a thermalized
configuration at sightly higher or lower temperature, respectively
(see, e.g.\ Ref.~\onlinecite{Nishimatsu_et_al_2008}). While an
appreciable thermal hysteresis is obtained for zero electric field,
the thermal hysteresis completely vanishes for fields above
20-30\,kV/cm. Therefore, only results from ``cooling'' runs are
presented in the following. In the vicinity of the phase transition,
due to the sharp features in $C_{p,\mathcal{E}}$, a dense 1\,K mesh
and extended equilibration time is used for field strengths below
75\,kV/cm. Otherwise, a temperature grid of 5\,K is used and the
specific heat is extrapolated to a 1\,K temperature grid and a moving
average is used to further smooth the data. Above 450\,K and for field
strengths of more than 200\,kV/cm, the total energy varies only
weakly.  Therefore, we have used a coarser temperature grid of 10\,K
in that region.

Using $\left( \partial P/\partial T\right)_\mathcal{E}$ and the
calculated $C_{p,\mathcal{E}}$, we can then obtain $\Delta T$ from
Eq.~\eqref{eq:deltaT-ind} and $\Delta S$ from
Eq.~\eqref{eq:deltaS-dir}. We note that we have confirmed the absence
of noticeable finite size effect in our results for electric fields
above $\sim$25\,kV/cm, which is above the critical field for the first
order phase transition. Therefore, using $\left(\partial P/\partial
T\right)_{\mathcal{E}}$ and $C_{p,\mathcal{E}}$ obtained from
different sizes of the simulation cell does not introduce any
significant errors or inconsistencies to our analysis.

In addition, we have performed test calculations assessing the effect
of different field ramping rates (see Sec.~\ref{sec:rates}). These
tests are performed using a $48 \times 48 \times 48$ simulation cell
and a time step of 1\,fs. Different thermalization and averaging times
have been used in these calculations, depending on the specific field
strength, ramping rate, and temperature. In all cases we verified that
the system is sufficiently equilibrated and averages were obtained
with good accuracy.

We note that in our calculations, we do not apply any empirical
pressure corrections, which have been used in previous studies to
correct for deficiencies of the first principles calculations or to
mimic thermal expansion. Such pressure corrections can lead to better
agreement between the calculated and measured transition
temperatures.~\cite{Nishimatsu_et_al_2010} However, as already pointed
out in Ref.~\onlinecite{Nishimatsu_Barr_Beckman:2013}, a
temperature-dependent pressure correction can also lead to
inconsistencies between the direct and indirect calculation of $\Delta
T$. Consequently, we refrain from using such pressure corrections (or
from rescaling the parameter $\kappa_2$ in the soft mode energy of the
effective Hamiltonian, see e.g.\ Ref.~\onlinecite{Walizer}) in this
work. As a result, our calculated transition temperature $T_c$ for the
cubic to tetragonal phase transition ($\sim270$\,K) deviates from the
known experimental value (403\,K).  However, it can be expected that
nevertheless trends are accurately described and that the calculated
temperature changes (after rescaling for the correct number of degrees
of freedom) are also quantitatively of the right magnitude.

\section{Results and Discussion}\label{sec:results}

\subsection{Rate dependence}
\label{sec:rates}

First, we investigate the influence of the rate of change,
$d\mathcal{E}/dt$, with which the electric field is ramped up or down
in our simulations. This is an important technical point, since,
depending of course on the invested computational resources, MD
simulations can only cover time periods of up to a few nano seconds.
This means that within the simulations, the electric field needs to be
changed extremely fast compared to a real experiment. Nevertheless, it
is very important to ensure that the system always stays in thermal
equilibrium and that the MD simulation indeed describes a reversible
process.

We start by analyzing the change of the total energy under application
of an electric field for the two cases of instantaneous electric field
switching and very slow ramping. In general, the change in total
energy $\Delta E_\text{tot}$ under application or removal of an
electric field is given by $\Delta E_\text{tot} = - \int P \cdot
d\mathcal{E}$, where $P$ is the polarization of the system. For
instantaneous switching, the polarization cannot follow the change of
the applied field and stays essentially constant during the switching
process. In the paraelectric phase, the spontaneous polarization is
zero. Therefore, when the field is switched on instantaneously for
$T>T_c$, $\Delta E_\text{tot}$ is also equal to zero. However, if the
field is instantaneously switched off from some finite value
$\mathcal{E}_\text{app}$, then even at $T > T_c$, there is an induced
polarization, $P_\text{ind} = \chi \mathcal{E}_\text{app}$, resulting
in a non-zero $\Delta E_\text{tot} = P_\text{ind} \cdot
\mathcal{E}_\text{app} = \chi \mathcal{E}_\text{app}^2$. This implies
that the complete cycle of applying and removing an electric field
instantaneously to the system results in an irreversible process.

\begin{table}
\renewcommand{\arraystretch}{1.5}
\centering
\begin{tabular}{lll}
\hline
 Case & Switching on & Switching off \\
\hline 
\multicolumn{3}{c}{Paraelectric phase ($T > T_c$)} \\
Instantaneous & 0 & $\chi\mathcal{E}_\text{app}^2$ \\
Ramping & $-\frac{1}{2}\chi\mathcal{E}_\text{app}^2$ & $\frac{1}{2}\chi\mathcal{E}_\text{app}^2$ \\
\hline
\multicolumn{3}{c}{Ferroelectric phase ($T < T_c$)} \\ 
Instantaneous & $-P_0\mathcal{E}_\text{app}$  & $P_0\mathcal{E}_\text{app} + \chi'\mathcal{E}_\text{app}^2$ \\
Ramping & $-P_0\mathcal{E}_\text{app} - \frac{1}{2}\chi'\mathcal{E}_\text{app}^2$ & 
$P_0\mathcal{E}_\text{app} + \frac{1}{2}\chi'\mathcal{E}_\text{app}^2$ \\
\hline
\end{tabular}
\caption{The changes in the total energy on varying the applied
  electric field are tabulated for instantaneous switching and slow
  ramping of the field.  ``Switching on'' corresponds to the field
  varying from zero to $\mathcal{E}_\text{app}$, and vice-versa for
  the ``switching off'' case.  The formulas are derived using the
  following simplified assumption: the induced polarization
  $P_\text{ind}$ depends linearly on the applied field $\mathcal{E}$,
  the proportionality constant is the dielectric susceptibility $\chi$
  in the paraelectric phase and $\chi'$ in the ferroelectric
  phase. $P_0$ is the spontaneous polarization of the system in the
  ferroelectric phase.}
\label{tab:dEtot-form}
\end{table}

On the other hand, if the field is applied/removed slowly, then the
polarization can follow the external field, and at each time
$P=\chi\mathcal{E}$. The resulting change in total energy is then
given according to $\Delta E_\text{tot} = -
\int_{\mathcal{E}_\text{i}}^{\mathcal{E}_\text{f}} P \cdot
d\mathcal{E} = \pm \tfrac{1}{2} \chi \mathcal{E}_\text{app}^2$. Here,
$\mathcal{E}_\text{i}$ and $\mathcal{E}_\text{f}$ are the initial and
final applied fields, respectively, which are equal to zero and
$\mathcal{E}_\text{app}$ for application of the field, and the other
way round for removal. The plus and minus signs then correspond to
removal and application of $\mathcal{E}_\text{app}$,
respectively. Thus, it can be seen that slow ramping results in the
same magnitude of the total energy change for switching the field on
and off, i.e.\ one obtains a reversible process.  Similar arguments
hold true within the ferroelectric phase, with an additional term
coming from the spontaneous polarization $P_0$. The resulting total
energy changes for the various cases are tabulated in
Table~\ref{tab:dEtot-form}.

\begin{figure}
\centering
\includegraphics*[width=0.8\columnwidth]{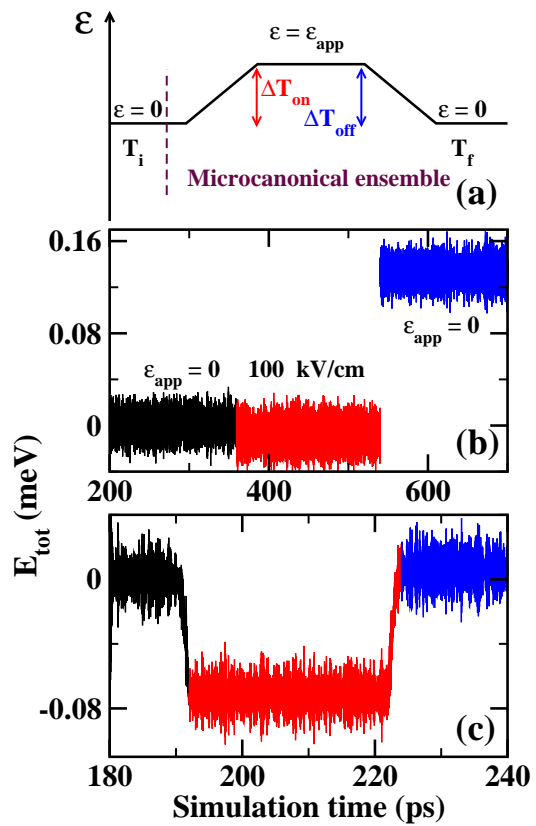}
\caption{(Color online) (a) Schematic depiction of the simulation
  cycle (see text). Panels (b) and (c) show the total energy as a
  function of MD steps for instantaneous field application/removal and
  for slow field ramping, respectively. Here, the starting temperature
  $T_i$ is 530\,K and the applied field is 100\,kV/cm. In (c) the rate
  of change of the applied field is equal to
  0.05\,kVcm$^{-1}$fs$^{-1}$.  For clarity, the total energy is
  plotted only after the system is switched to the microcanonical
  ensemble.}
\label{fig:Etot_field}
\end{figure}

Next, we perform simulations at different temperatures to examine
whether the simple considerations outlined in the preceding paragraphs
are consistent with the actual MD simulations for the effective
Hamiltonian. We have selected two temperatures, $T =$ 530\,K (in the
paraelectric phase) and $T =$ 270\,K (in the ferroelectric phase, just
below the transition temperature), at which we monitor the change in
the total energy on switching the field on and then off again for
several values of $\mathcal{E}_\text{app}$. The full simulation cycle
is depicted schematically in Fig.~\ref{fig:Etot_field}(a). First, the
system is thermalized at temperature $T_\text{i}$ and field
$\mathcal{E}=0$ within the canonical ensemble. The simulation is then
switched to the microcanonical ensemble, the electric field is ramped
up to $\mathcal{E}_\text{app}$, and the resulting temperature change
$\Delta T_\text{on}$ is monitored. Then, the field is ramped down
again, and the corresponding temperature change $\Delta T_\text{off}$
is monitored. If the system stays in thermal equilibrium during the
entire simulation cycle, then both its total energy and its
temperature, $T_\text{f}$, at the end of the simulation should be
identical to the corresponding starting values, and $\Delta
T_\text{off} = - \Delta T_\text{on}$.

The evolution of the total energy over a full cycle at
$T_\text{i}=530$\,K, i.e.\ in the paraelectric phase, is shown in
Fig.~\ref{fig:Etot_field}(b). In this simulation, the field is
switched instantaneously. As expected, there is no change in the total
energy while switching on the field (see Table~\ref{tab:dEtot-form}),
but there is a jump of the total energy when the field is switched
off. Fig.~\ref{fig:Etot_field}(c) shows the evolution of the total
energy when the field is ramped up and down slowly. In this case,
$|\Delta E_\text{tot}|$ is the same for application and removal of the
field. This confirms that very fast switching of the applied field
results in an irreversible process.

\begin{figure}[tb]
\includegraphics[width=0.7\columnwidth]{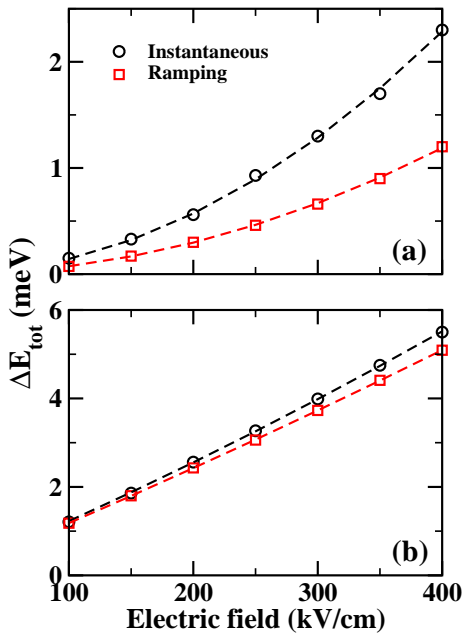}
\caption{(Color online) The change in the total energy on switching
  off the applied field is plotted as a function of applied electric
  field for (a) the paraelectric phase at $T_i =$ 530\,K and (b) the
  ferroelectric phase at $T_i =$ 270\,K.  The rate of change of the
  applied field $d\mathcal{E}/dt$ is equal to
  $-$0.05\,kVcm$^{-1}$fs$^{-1}$ for the ``ramping'' case (red
  squares). The symbols show the data points, whereas the dashed lines
  are fits to the corresponding functional forms listed in
  Table~\ref{tab:dEtot-form}.}
\label{fig:delEtot_field}
\end{figure}

Further, we plot the change in the total energy $\Delta E_\text{tot}$
as a function of the applied field $\mathcal{E}_\text{app}$ at
$T_\text{i} =$ 530\,K and $T_\text{i} =$ 270\,K in
Figs.~\ref{fig:delEtot_field}(a) and (b), respectively. These $\Delta
E_\text{tot}$ values correspond to removal of the field (``switching
off'').  For slow ramping, these are equal to those obtained from
switching on (but with opposite sign). The data from the simulations
is fitted using the corresponding functional forms given in
Table~\ref{tab:dEtot-form}. The fits are indicated by dashed lines and
match very well with the data. In the paraelectric phase the total
energy change for switching off the field depends quadratically on the
applied field strength and there is a factor of 2 difference between
slow ramping and instantaneous switching. The fit to the instantaneous
switching data gives $\chi = 3.5 \times 10^{-2} \, \mu\text{C} \cdot
\text{kV}^{-1} \text{cm}^{-1}$ which matches well with the
corresponding value obtained from the ramping data ($\chi = 3.8 \times
10^{-2} \mu\text{C} \cdot \text{kV}^{-1} \text{cm}^{-1}$).  In the
ferroelectric phase, $\Delta E_\text{tot}$ is dominated by the linear
contribution stemming from the spontaneous polarization. From the fit
of the instantaneous switching data, we obtain
$P_0=30.32$\,$\mu$C/cm$^2$ and $\chi'= 1.3 \times
10^{-2}$\,$\mu$C$\cdot$kV$^{-1}$cm$^{-1}$, whereas for the case of
slow ramping the corresponding quantities are 30.32\,$\mu$C/cm$^2$ and
$1.5 \times 10^{-2}$\,$\mu$C$\cdot$kV$^{-1}$cm$^{-1}$
respectively. There is good agreement between these two data
sets. Similarly, the value for $P_0$ obtained from fitting $\Delta
E_\text{tot}$ for instantaneous ``switching on'' of the electric field
(not shown here) is equal to 30.31\,$\mu$C/cm$^2$. We can also compare
these values for the spontaneous polarization to that obtained
directly from the MD simulations at $T =$ 270\,K, which is equal to
28.2\,$\mu$C/cm$^2$.  Note that the agreement between the various
parameters is excellent considering the simplicity of the
approach. This shows that the simple considerations outlined at the
beginning of this section do indeed lead to a consistent description
of the various switching cases.

\begin{figure}[tb]
\includegraphics[width=\columnwidth]{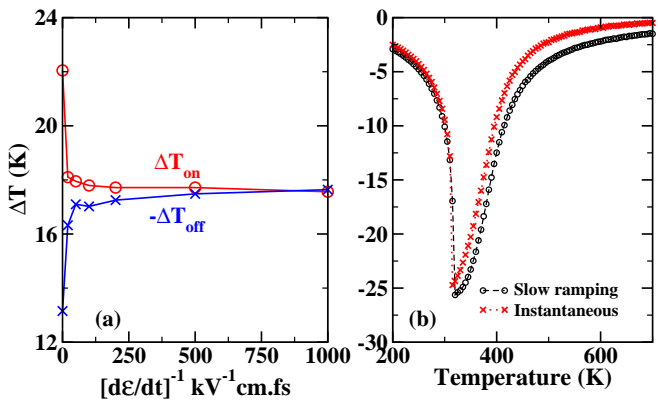}
\caption{(Color online) (a) EC temperature changes for switching the
  electric field on and off, $\Delta T_\text{on}$ and $\Delta
  T_\text{off}$, as function of the inverse rate of change of the
  applied electric field, $(d\mathcal{E}/dt)^{-1}$, for a starting
  temperature $T_\text{i} = 350$\,K. The applied field
  $\mathcal{E}_\text{app}$ is equal to 200\,kV/cm. The simulation
  cycle used to obtain $\Delta T_\text{on/off}$ is depicted in
  Fig.~\ref{fig:Etot_field}(a). (b) The EC temperature change
  $\Delta T$ as a function of temperature is plotted for slow ramping
  and instantaneous ramping of the field. The field is varied from
  225\,kV/cm to 0\,kV/cm.}
\label{fig:deltaT-ratedep}
\end{figure}

Fig.~\ref{fig:deltaT-ratedep}(a) shows the obtained temperature
changes when the electric field is switched on and off, $\Delta
T_\text{on}$ and $\Delta T_\text{off}$, as function of the inverse
rate $\left(d\mathcal{E}/dt \right)^{-1}$, with which the field is
ramped up and down, for a starting temperature $T_\text{i} = 350$\,K
(i.e.\ close to the maximum EC effect). Instantaneous switching
corresponds to $(d\mathcal{E}/dt)^{-1} = 0$ and slow ramping
corresponds to a large inverse rate. It can be seen that for small
inverse ramping rates, i.e.\ for fast switching, $|\Delta T_\text{on}|
\ne |\Delta T_\text{off}|$.  The difference between the values is
about 9\,K which is significant compared to the converged EC
temperature change of 17.6\,K observed at this
temperature.~\cite{comment} This implies that the system goes out of
equilibrium during fast switching, consistent with the total energy
considerations discussed above. As the rate is reduced (i.e.\ the
inverse rate is increased), the difference between $|\Delta
T_\text{on}|$ and $|\Delta T_\text{off}|$ becomes smaller, and already
for an inverse rate of 500\, kV$^{-1}$$\cdot$fs$\cdot$cm
(corresponding to $|d\mathcal{E}/dt| = 0.002$\,kVcm$^{-1}$fs$^{-1}$),
the difference becomes negligible. Similar behavior can be observed
also for other initial temperatures.

In Fig.~\ref{fig:deltaT-ratedep}(b), the directly calculated EC
temperature change obtained by instantaneously switching off the
electric field is compared with the one obtained using slow field
ramping ($d\mathcal{E}/dt = -0.002\,\text{kV} \cdot \text{cm}^{-1}
\text{fs}^{-1}$) for an initially applied field of 225\,kV/cm. Note
that for these calculations the system is thermalized in the canonical
ensemble with a nonzero field, which is removed after switching to the
microcanonical ensemble. While the overall behavior is the same for
both cases, the instantaneous switching underestimates the magnitude
of $\Delta T$ at essentially all temperatures. This is consistent with
our earlier observations in Fig.~\ref{fig:deltaT-ratedep}(a). In
addition, the temperature for which the largest temperature change
occurs is slightly shifted to lower temperatures.

The maximum $\Delta T$, observed at 320\,K in
Fig.~\ref{fig:deltaT-ratedep}(b), is about $-$25.6\,K for slow
ramping. After appropriate rescaling for the correct number of degrees
of freedom,\cite{Nishimatsu_Barr_Beckman:2013,comment} this corresponds
to an EC temperature change of around 5.1\,K, which agrees well with
earlier reports for similar applied field
strengths.~\cite{Beckman_et_al_2012,Marathe/Ederer:2014} After
scaling, the difference between $\Delta T$ calculated using
instantaneous switching and using slow field ramping is approximately
1\,K for the given field strength (except very close to the peaks).

Our results up to now thus demonstrate the necessity of ensuring that
the system is in thermal equilibrium throughout the whole MD
simulation for a correct direct calculation of the adiabatic EC
temperature change. In the following we use a rate $d\mathcal{E}/dt =
0.002$\,kVcm$^{-1}$fs$^{-1}$ for all our direct calculations of the EC
temperature change. Even though this rate is very fast compared to
actual experimental rates, it is sufficiently slow to avoid
irreversibility in the calculation and also allows for reasonable
simulation times.

\subsection{Direct versus indirect EC effect}
\label{sec:dir-indir}

Next, we compare results from the direct and indirect approaches.  As
pointed out previously, all calculations are performed using the same
effective model Hamiltonian and no experimental data is used. In our
previous work, we have calculated the EC effect using the indirect
method,~\cite{Marathe/Ederer:2014,Beckman_et_al_2012} but we have used
the experimental specific heat value at room temperature to evaluate
Eq.~\eqref{eq:deltaT-ind}.  Although this is a valid first
approximation, this treatment ignores the temperature and electric
field dependence of $C_{p,\mathcal{E}}$, as well as the mismatch
between the number of degrees of freedom of the real system and the
model, which will lead to differences between the results obtained
from direct and indirect methods. Therefore, in the present work, we
calculate the specific heat from the model Hamiltonian as a function
of temperature at different applied fields. This allows for an
internally consistent comparison between direct and indirect methods,
and also enables us to obtain $\Delta S^{\text{dir}}$ using the
(quasi-) direct method, Eq.~\eqref{eq:deltaS-dir}.

\begin{figure}[tb]
\centerline{\includegraphics[width=0.9\columnwidth]{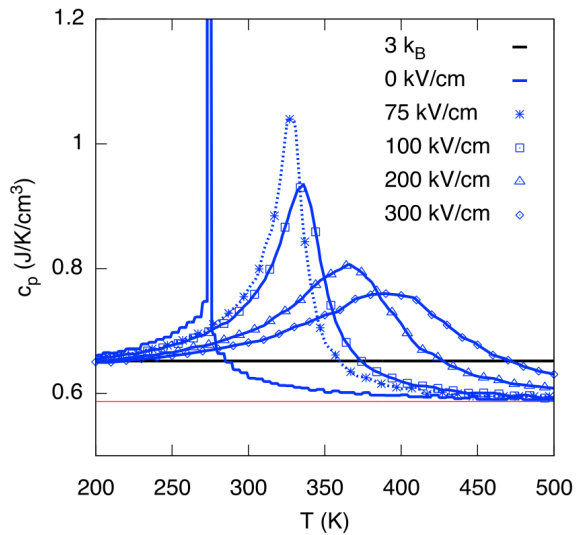}}
\caption{(Color online) Calculated specific heat of the model
  Hamiltonian as a function of temperature for different applied
  electric fields. For better comparison, the Dulong-Petit value of 3
  $k_B$ is indicated by the thick horizontal black line and the
  obtained high temperature limit is indicated by the thin horizontal
  red line.}
\label{fig:cppaper}
\end{figure}

In Fig.~\ref{fig:cppaper}, we show our results for the specific heat
of the effective Hamiltonian as a function of temperature at several
applied fields. In absence of an applied field, there is a pronounced
peak (divergence) at the ferroelectric transition, which also shows
pronounced thermal hysteresis.  Such a divergence is characteristic
for a first order phase transition.  With increasing electric field,
the phase transition and thus the peak in the specific heat shift to
higher temperature.  Furthermore, the transition becomes smoother and
the thermal hysteresis disappears for fields of around 25\,kV/cm and
stronger. Previous phenomenological thermodynamic calculations and
experimental measurements on bulk BaTiO$_3$ have shown that an applied
field of about 10\,kV/cm is sufficient to suppress the first order
phase
transition.~\cite{Zhang_et_al_2009,Novak,Novak_Pirc_Kutnjak_2013}

Both at high and low temperatures, i.e.\ away from the phase
transition, $C_{p,\mathcal{E}}$ approaches constant values.  We recall
that our simulations contain only 3 degrees of freedom instead of 15
for the real system. If the Hamiltonian would be exactly quadratic in
the 3 soft mode variables, then each degree of freedom would
contribute $1k_B$ to the specific heat, and we would expect a value of
3$k_B$, which is equal to 0.65\,J$\cdot$K$^{-1}$cm$^{-3}$, both in the
high and low temperature limit. We note that in our purely classical
simulations no modes are ``frozen in'' at low temperatures. We find a
high-temperature limit of 0.59\,J/K/cm$^3$ from our calculations. This
small discrepancy with the Dulong-Petit value of $3k_{B}$ results from
the higher order terms in the Hamiltonian. These contribute less at
low temperatures, and therefore at low temperatures below $T_c$ the
calculated value compares well with the Dulong-Petit value.

\begin{figure}[tb]
\includegraphics[width=\columnwidth]{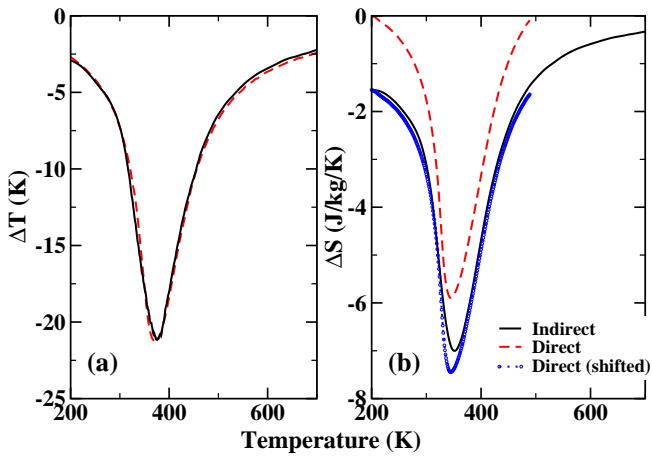}
\caption{(Color online) (a) Comparison of the EC temperature change as
  function of the initial temperature obtained using direct and
  indirect methods. The field is varied from 300\,kV/cm to 75\,kV/cm.
  In the direct calculation slow field ramping with $d\mathcal{E}/dt =
  -0.002$\,kVcm$^{-1}$fs$^{-1}$ is used. (b) The EC entropy change as
  a function of temperature obtained from direct and indirect
  methods. Here, the applied field is varied from 75\,kV/cm to
  300\,kV/cm. The results of the direct calculations, but with an
  offset such that it matches the indirect calculation at $T=200$\,K
  (see text) are indicated by the blue dotted line.}
\label{fig:deltaT-comp}
\end{figure}

Next, we compare both the calculated adiabatic EC temperature change
$\Delta T$ as well as the isothermal EC entropy change $\Delta S$
obtained using direct and indirect methods. As mentioned before, we
use a rate of $-$0.002\,kVcm$^{-1}$fs$^{-1}$ for changing the applied
field. In Fig.~\ref{fig:deltaT-comp}(a), we compare the results from
direct and indirect calculations of the adiabatic EC temperature
change $\Delta T$. These results correspond to removal of the field
(``switching off''), i.e.\ a negative $\Delta T$ (as in
Fig.~\ref{fig:deltaT-ratedep}(b)). We consider only fields $\geq$
75\,kV/cm for this comparison, in order to exclude the region close to
the first order phase transition, where the indirect method is not
applicable. It can be seen that the $\Delta T$ values calculated from
direct and indirect methods match extremely well over the whole
temperature range. This shows that within our consistent description,
where all quantities are calculated using the same effective
Hamiltonian, the direct and indirect approaches indeed lead to exactly
the same $\Delta T$. This underlines the validity of the indirect
approach to obtain $\Delta T$, which is often preferred in experimental
studies, as long as the system does not actually cross the first order
phase transition, i.e.\ for temperatures and electric field strengths
above the critical point.~\cite{Novak,Novak_Pirc_Kutnjak_2013} We
point out, though, that errors can be introduced due to imperfect fits
to the statistical measurements, in particular for small fields close
to the phase transition, where polarization and specific heat vary
strongly. Another source of inaccuracies is the finite sampling of
quantities as function of temperature and field, which leads to
numerical errors when integrating Eq.~\eqref{eq:deltaT-ind}.

\begin{figure}[tb]
\centerline{\includegraphics*[width=0.85\columnwidth]{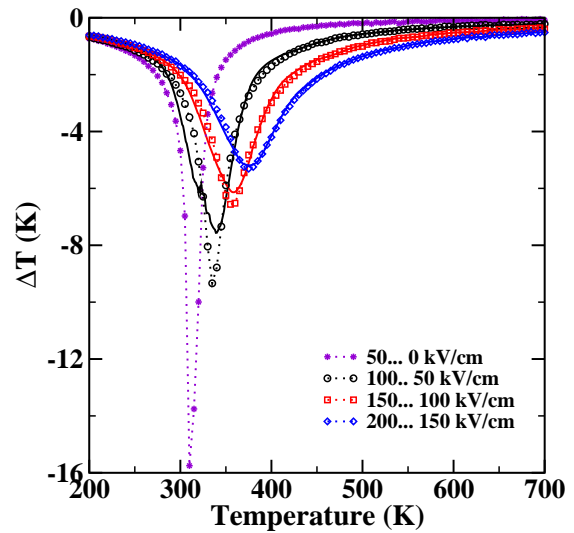}}
\caption{(Color online) EC temperature change $\Delta T$ as function
  of temperature for different field intervals with the same total
  width of 50\,kV/cm. Results obtained using the direct (indirect)
  method are shown as dotted lines with symbols (solid lines without
  symbols).}
\label{fig:deltaT-order}
\end{figure}

The isothermal entropy change calculated using direct
(Eq.~\eqref{eq:deltaS-dir}) and indirect (Eq.~\eqref{eq:deltaS-ind})
methods is shown in Fig.~\ref{fig:deltaT-comp}(b). We note that in
principle the specific heat in Eq.~\eqref{eq:deltaS-dir} is the total
specific heat, including electronic, ferroelectric, and all other
lattice contributions, while in our treatment using the effective
Hamiltonian, only the contributions from the ferroelectric soft mode
variables are taken into account. However, since we use exactly the
same degrees of freedom to also obtain the temperature- and
field-dependent electric polarization for evaluating
Eq.~\eqref{eq:deltaS-ind}, we obtain a consistent description within
the effective Hamiltonian, and both equations should in principle lead
to the same value of $\Delta S$.  Nevertheless, in contrast to the
adiabatic temperature changes, the calculated isothermal entropy
changes do not need to be rescaled for the missing degrees of freedom
in order to compare with experimental measurements. The entropy change
in Eq.~\eqref{eq:deltaS-dir} depends only on differences in the
specific heat for different electric fields, and one can expect, at
least to a good approximation, that only the soft mode variables will
give significant contributions to this electric field
dependence. Consequently, also Eq.~\eqref{eq:deltaS-ind} does not
contain any quantities depending explicitly on the missing degrees of
freedom. The effect of the electrons and of other structural degrees
of freedom on the temperature and field dependence of the polarization
is, to a good approximation, implicitly taken into account by the soft
mode variable.

The entropy change shown in Fig.~\ref{fig:deltaT-comp}(b) corresponds
to a change of the electric field from 75\,kV/cm to 300\,kV/cm,
i.e.\ ``switching on'', and therefore a negative $\Delta S$ is
obtained. Again, we exclude the region of small electric fields close
to the phase transition for the comparison of direct and indirect
methods. The lower bound for the temperature integration in
Eq.~\eqref{eq:deltaS-dir} is chosen as $T_1$ = 200\,K. This
temperature is above the second phase transition from the tetragonal
ferroelectric to the orthorhombic ferroelectric phase in BaTiO$_3$
(with the chosen parameterization of the effective Hamiltonian). By
definition, the ``direct'' $\Delta S$ calculated from
Eq.~\eqref{eq:deltaS-dir} is zero for $T=T_1$, while the ``indirect''
$\Delta S$ obtained from Eq.~\eqref{eq:deltaS-ind} has a finite value
at $T_1=200$\,K. Since according to Fig.~\ref{fig:cppaper} the
calculated specific heat at 200\,K shows only negligible field
dependence, the finite value of $\Delta S$ at this temperature is
related to electric-field dependence of the specific heat at lower
temperatures, most likely at the two ferroelectric-ferroelectric
transitions (tetragonal-orthorhombic and
orthorhombic-rhombohedral). For a better comparison between direct and
indirect methods we therefore rigidly shift the $\Delta S$ curve
obtained from the direct method such that it matches the $\Delta S$
value obtained from the indirect method at the lowest temperature
$T_1=200$\,K (blue dotted line in Fig.~\ref{fig:deltaT-comp}). It can
be seen that the shifted data agrees quite well with the data obtained
from the indirect method. Small deviations can be observed close to
the peak at around 350\,K, which we ascribe to inaccuracies related to
the smoothing/fitting of the specific heat and polarization data and
to integration errors due to finite temperature and electric field
sampling.

Interestingly, the obtained peak value of \mbox{$\Delta S \approx
  7$\,J$\cdot$kg$^{-1}$K$^{-1}$} is of the same order of magnitude as
the maximal value reported for BaTiO$_3$ single crystals measured in
Ref.~\onlinecite{Moya_et_al:2013} ($\Delta S =
2.1$\,J$\cdot$kg$^{-1}$K$^{-1}$). However, the corresponding electric
field intervals are completely different (75 to 300\,kV/cm in our
calculations, compared to 0 to 4\,kV/cm in
Ref.~\onlinecite{Moya_et_al:2013}), and thus a meaningful quantitative
comparison is not easily possible. 

Finally, in Fig.~\ref{fig:deltaT-order} we compare the calculated
adiabatic temperature change for different electric field intervals
with the same width $|\mathcal{E}_\text{i} -
\mathcal{E}_\text{f}|=50$\,kV/cm but different magnitude of
$\mathcal{E}_\text{i}$ and $\mathcal{E}_\text{f}$. A first order phase
transition occurs only in the electric field interval between $\mathcal{E}_\text{i} =
50$\,kV/cm and $\mathcal{E}_\text{f}=0$\,kV/cm. For this interval, we
obtain a very narrow peak in $\Delta T$ at 310\,K, with a maximum
value of 15.8\,K (corresponding to $\sim$3.2\,K after scaling to the
correct $N_f$). For larger applied fields, the maximum $\Delta T$
value shifts to higher temperatures and the corresponding peak
broadens.

We note that from our specific heat calculations we can estimate a
critical electric field of around 25\,kV/cm. For larger fields, the
polarization varies continuously with temperature, i.e.\ no first order
phase transition occurs, and the temperature at which $\left|\partial
P/\partial T \right|_\mathcal{E}$ is maximal follows the so-called
``Widom line'' (see
e.g.\ Refs.~\onlinecite{Novak_Pirc_Kutnjak_2013,Rose/Cohen:2012}). It
can be seen that, while the largest $\Delta T$ is observed in the
field interval containing the first order phase transition
(i.e.\ $\mathcal{E}_\text{i}=50$\,kV/cm and $\mathcal{E}_\text{f} =
0$\,kV/cm), the field intervals corresponding to larger
$\mathcal{E}_\text{i}$/$\mathcal{E}_\text{f}$ also give sizable
contributions to the EC effect. We therefore conclude that while the
vicinity to the first order transition is important to obtain large
changes of polarization with temperature and electric field, and thus
large EC effect, the contribution of the transition itself is not
essential to obtain large EC response. We note that similar
conclusions have been reached in Ref.~\onlinecite{Rose/Cohen:2012},
based on MD simulations for LiNbO$_3$.

Further, a comparison between $\Delta T$ obtained using the direct and
indirect methods for the different field intervals again shows a very
good agreement for the field intervals corresponding to larger
magnitude of $\mathcal{E}_\text{i}$ and $\mathcal{E}_\text{f}$, where
the variation with temperature and electric field is less
strong. Clear discrepancies can be seen for the interval between 100
and 50\, kV/cm, where $\Delta T$ is rather sharply peaked. These
discrepancies result from imperfect smoothing/fitting as well as from
numerical integration errors, as already discussed above.

\section{Summary and Conclusions}

In summary, we have presented a computational study of the EC effect
in BaTiO$_3$ using MD for a first principles-based effective
Hamiltonian. We have compared the EC temperature change calculated
using direct and indirect methods for bulk BaTiO$_3$, thereby paying
particular attention to the internal consistency of the method. In
particular, the temperature and electric field-dependent specific heat
has been calculated within the same framework as the temperature- and
field-dependent electric polarization (required for the indirect
determinations of $\Delta T$), and the same framework has also been
used for the direct calculation of $\Delta T$ using microcanonical MD.

We have demonstrated that the direct and indirect determination of the
adiabatic temperature change leads to identical results provided that
the field and temperature region very close to the first order
transition, where the indirect method is not applicable, is
excluded. We note that the applicability of Maxwell's relation,
Eq.~\eqref{eq:maxwell}, which underlies the indirect determination of
the EC temperature change, has been critically discussed for systems
close to a first order phase transition (see e.g.\ the discussion in
Refs.~\onlinecite{Valant_2012,ECE-Book,Moya/Narayan/Mathur:2014}). Our
results clearly demonstrate the validity of this relation as long as
the first order transition is not crossed. Directly at the first order
transition, the specific heat diverges and $\partial P/\partial T$ is
not defined, and thus the indirect method is not applicable. Very
close to the transition, errors can arise due to inaccurate fits, the
use of a temperature- and field-independent specific heat, and due to
finite temperature and field sampling of the integral in
Eq.~\eqref{eq:deltaT-ind}.

Furthermore, we have demonstrated the importance of maintaining
thermal equilibrium during the MD simulations for the direct
calculatison for $\Delta T$, and we have shown that in the present
case a ramping rate for the electric field of 0.002\,kV/cm/fs is
sufficiently slow to ensure reversibility. We note, however, that, due
to the neglect of the less important degrees of freedom in the
effective Hamiltonian, this is not necessarily representative for the
intrinsic relaxation time of BaTiO$_3$.

In addition, we have (to the best of our knowledge for the first time)
used the effective Hamiltonian approach to calculate the isothermal EC
entropy change. Similarly to the case of the adiabatic temperature
change, we have found good agreement between (quasi-) direct, i.e.\ via
the specific heat, and indirect determination of $\Delta S$. While our
calculated values are quantitatively of similar magnitude as available
experimental data, further studies for different electric field
strengths, possibly also considering the contribution stemming from
the latent heat of the first order phase transition, are necessary to
obtain a more quantitative comparison between calculated and measured
data.

The observation that the largest EC temperature change occurs in the
field interval containing the first order ferroelectric transition
($\mathcal{E}_\text{i}=50$\,kV/cm and $\mathcal{E}_\text{f}=0$\,kV/cm,
see Fig.~\ref{fig:deltaT-order}) is in agreement to the giant caloric
temperature changes found at other coupled ferroic-structural
transitions, e.g.\ in magnetic Heusler
alloys.~\cite{Planes,Liu2,Moya/Narayan/Mathur:2014} In all these
cases, small external fields are able to induce large adiabatic
temperature changes, which, however, are restricted to only a narrow
temperature interval. Unfortunately, in many cases the thermal
hysteresis of the transition leads to a significant reduction of the
achievable reversible temperature changes under cycling of the fields,
see e.g.\ Ref.~\onlinecite{Liu2}. In this respect, the EC effect in
BaTiO$_3$, with its rather low critical field strength, has an
important advantage for cooling applications compared to the
well-established magnetocaloric Heusler alloys. Beyond the critical
field strength, the thermal hysteresis vanishes, leading to a fully
reversible EC effect (see also the experimental results in
Ref.~\onlinecite{KarNarayan}). Furthermore, the transition itself is
not crucial for obtaining a large caloric response, as field intervals
corresponding to larger $\mathcal{E}_\text{i}$/$\mathcal{E}_\text{f}$,
i.e.\  above the critical field strength, also give sizable
contributions to the EC effect, see Fig.~\ref{fig:deltaT-order}. This
is accompanied by a broadening of the $\Delta T$ peak with
temperature, which is also advantageous for applications. In contrast,
the strong first order character of the magneto-structural phase
transition in magnetic Heusler alloys is conserved even for giant
fields up to 40\,T, without a reduction of the thermal
hysteresis.~\cite{Mejia}

\section{Acknowledgments}
This work was supported by the Swiss National Science Foundation and
the German Science Foundation under the priority program SPP 1599
(``Ferroic cooling'').  AG thanks the CCSS at the University of
Duisburg-Essen for computing time. The work of TN was supported in
part by JSPS KAKENHI Grant Number 25400314.

\bibliography{references}

\end{document}